\documentclass[usenatbib]{mn2e}

\pdfoutput=1

\usepackage{fixltx2e}
\usepackage{mathptmx}
\usepackage[pdftex]{graphicx}

\newcommand{\Mpc}{\rm\thinspace Mpc}

\newcommand{\km}{\rm\thinspace km}

\newcommand{\cm}{\rm\thinspace cm}
\newcommand{\yr}{\rm\thinspace yr}

\newcommand{\s}{\rm\thinspace s}

\newcommand{\Msun}{\hbox{$\rm\thinspace M_{\odot}$}}

\newcommand{\Msunpyr}{\hbox{$\Msun\yr^{-1}\,$}}

\newcommand{\kmps}{\hbox{$\km\s^{-1}\,$}}

\newcommand{\kmpspMpc}{\hbox{$\kmps\Mpc^{-1}$}}

\newcommand{\Zsun}{\hbox{$\thinspace \mathrm{Z}_{\odot}$}}

\newcommand{\psqcm}{\hbox{$\cm^{-2}\,$}}
\newcommand{\pcmsq}{\hbox{$\cm^{-2}\,$}}

\begin{document}

\title[O\,\textsc{vii} from \emph{XMM-Newton} RGS] {Revealing
  O\,\textsc{vii} from stacked X-ray grating spectra of clusters,
  groups and elliptical galaxies}

\author
[J.S. Sanders et al]
{J.~S. Sanders and A.~C. Fabian
  \\
  Institute of Astronomy, Madingley Road, Cambridge. CB3 0HA\\
}
\maketitle

\begin{abstract}
  We stack 4.6\,Ms of high spectral resolution \emph{XMM-Newton}
  Reflection Grating Spectrometer spectra from galaxy clusters, groups
  of galaxies and elliptical galaxies. For those objects with a
  central temperature of less than 1~keV, we detect O\,\textsc{vii}
  for the first time, with a probability of false detection of
  $2.5\times10^{-4}$. The flux ratio of the O\,\textsc{vii} to
  Fe\,\textsc{xvii} lines is $1/4$ to $1/8$ of the emission expected
  for isobaric radiative cooling in the absence of heating. There is
  either a process preventing cooling below 0.5~keV, anomalous O/Fe
  abundance ratios, absorbing material around the coolest X-ray
  emitting gas or non-radiative cooling taking place. The mean
  N\,\textsc{vii} emission line is strong in the sub-keV sample. As
  the ratio of the hydrogenic N and O lines is largely independent of
  temperature, we measure a mean N/O ratio of $4.0 \pm 0.6
  \Zsun$. Although the continuum around the C\,\textsc{vi} lines is
  difficult to measure we can similarly estimate that the C/O ratio is
  $0.9 \pm 0.3 \Zsun$.
\end{abstract}


\section{Introduction}
The intracluster medium (ICM) is the few $\times10^{6}$ to $\sim
10^8$~K X-ray emitting plasma which fills the potential wells of
objects ranging in scale from galaxy clusters down to massive
elliptical galaxies. The X-ray surface brightness profile of these
objects is often steeply peaked \citep[e.g.][]{Stewart84}. The implied
mean radiative cooling times are less than 1~Gyr. In the absence of
heating 10s to 100s of solar masses per year should be cooling below
$\sim10^{6}$K to form a cooling flow \citep{Fabian94}.

The relative strength of X-ray emission lines is a good probe of the
temperature distribution of the ICM. The major spectral lines in the
$5-38${\AA} spectral range can be observed in bright objects using the
Reflection Grating Spectrometer (RGS) instruments on
\emph{XMM-Newton}.

The weak or missing Fe\,\textsc{xvii} emission lines, indicating
material around $\sim 0.5$~keV, in RGS spectra showed that there is
much less cool gas present in clusters than expected from cooling in
the absence of heating \citep[e.g.][]{Peterson01,Peterson03}. However,
these lines have been detected in several objects using deep
observations \citep{SandersRGS08,SandersA220409}. There can be a wide
range of $\times10-15$ in temperature, but less than 10 per cent of
the emission expected for radiative cooling is seen.  Active galactic
nuclei (AGN) in the cores of clusters are believed to be the mechanism
by which energy is supplied to the ICM to combat a large fraction of
the radiative cooling \citep[for reviews
see][]{PetersonFabian06,McNamaraNulsen07}.

\begin{figure}
  \includegraphics[width=\columnwidth]{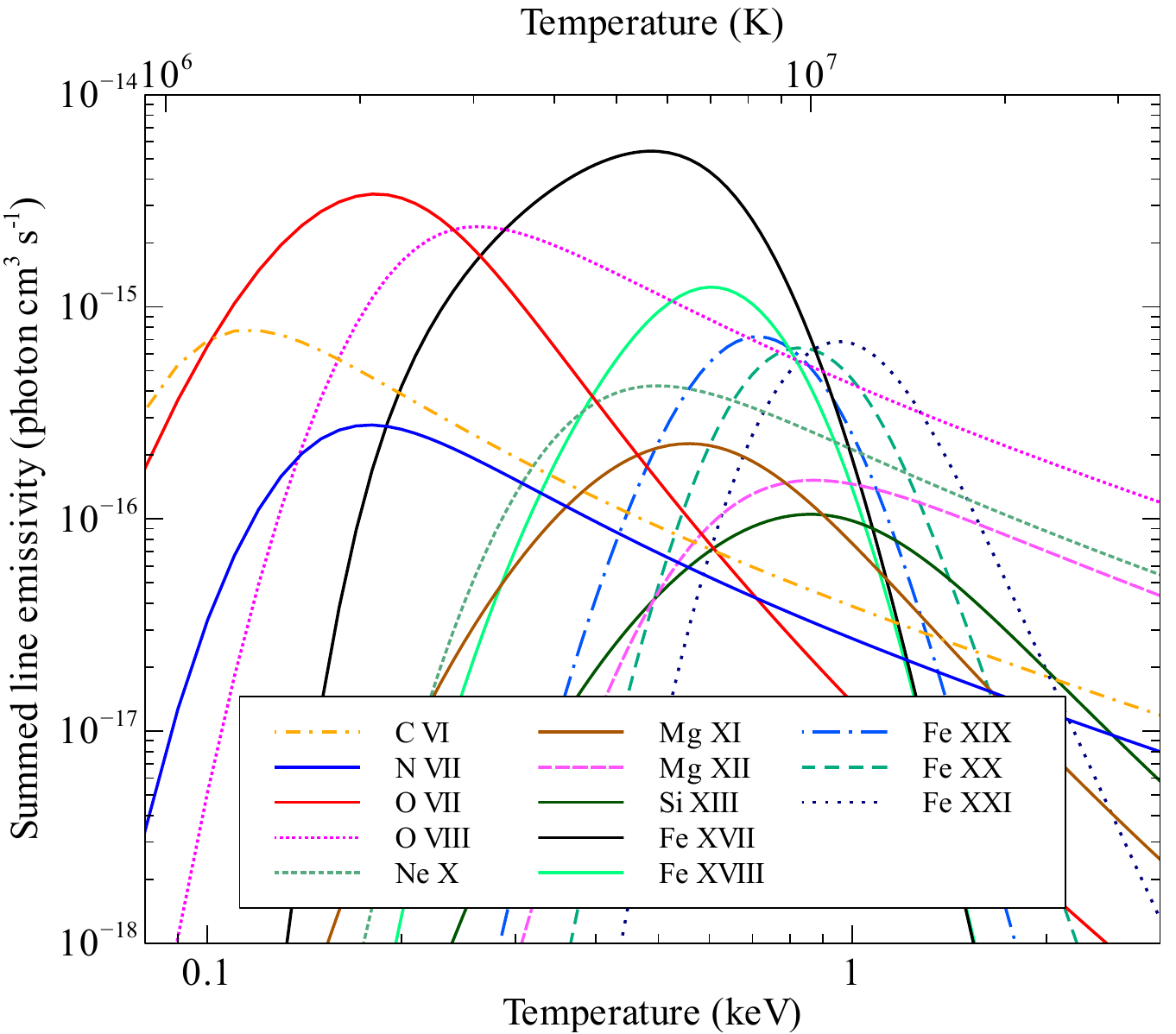}
  \caption{Emissivity of lines in the RGS waveband from various
    strongly X-ray emitting ions as a function of temperature at Solar
    metallicity.}
  \label{fig:linestrengths}
\end{figure}

Fig.~\ref{fig:linestrengths} shows the emissivity of emission lines
from certain strong X-ray emitting ions as a function of
temperature. To create this plot we summed the emissivity of strong
lines in the RGS spectral range calculated using \textsc{spex} 2.00.11
\citep{Kaastra00}. The emissivity in Fe\,\textsc{xvii} emission peaks
at a temperature of around 0.5~keV. To observe gas at lower
temperatures requires the detection of O\,\textsc{vii} lines. No
significant detections of O\,\textsc{vii} have been reported in
clusters, groups or elliptical galaxies, implying there is little
material below 0.5~keV temperature.

Very deep \emph{XMM} observations of bright high-metallicity cool
objects may detect the presence of O\,\textsc{vii} emission lines. The
calorimeters planned for \emph{ASTRO-H} and \emph{IXO} should also be
able to detect these lines. In the absence of any such observations at
this time we conducted a stacking analysis of the best observations in
the \emph{XMM-Newton} archive.

In this paper we assume the relative Solar abundances of
\cite{AndersGrevesse89}.

\section{Spectral stacking}
We took the processed RGS spectra from our analysis of the line
broadening in 62 galaxy clusters, groups and elliptical galaxies
\citep{Sanders10_Broaden}. These spectra were extracted from 90 per
cent of the point spread function (PSF), a roughly 50~arcsec wide
strip across the centre of the object, and included 90 per cent of the
pulse height distribution. We removed from the sample the two objects,
Klemola 44 and RX\,J1347.5-114, which showed significant line
broadening.

We note that the sample is not complete and was originally chosen to
contain objects with strong emission lines and a peaked surface
brightness profile. The results are likely to be relevant for this
sort of object, but perhaps not the average cluster, group or
elliptical galaxy.

For each object we created a fluxed spectrum with \textsc{rgsfluxer},
combining the two first order and two second order spectra. A fluxed
spectrum uses a fixed wavelength binning and is divided by the
effective area of the telescope to be in units of
photon$\cm^{-2}\s^{-1}$\AA$^{-1}$. We used a bin size of 0.01{\AA}
when fluxing the spectra. We background-subtracted the spectra using a
background extracted from beyond 98 per cent of the RGS PSF \citep[as
in][]{Sanders10_Broaden}. The \textsc{rgsfluxer} tool is not normally
used in the quantitative analysis of spectra because it does not take
account of the redistribution into neighbouring channels. For the low
resolution spectra we examine here and because we only fit the
emission lines with simple Gaussian functions, this is not a serious
issue.

The wavelength bins of the fluxed spectra were corrected for redshift
(note that we did not change the flux, which is still the flux in the
observed frame). We rebinned and added these spectra to make a stacked
spectrum with 0.04{\AA} resolution. For each output bin we divided by
the total number of input bins and spectra which were added for that
bin, to construct an average spectrum.

\begin{table}
  \caption{Details of the objects in the different temperature
    bins. Shown are the number of objects in the bin, the average
    redshift and the total RGS1 plus RGS2 exposures.}
  \begin{tabular}{llcl}
    \hline
    Temperature range (keV) & Number & $<z>$ & Exposure (ks) \\
    \hline
    $0-0.7$ & 9  & 0.0063 & 1240 \\
    $0.7-1$ & 9  & 0.012  & 1329 \\
    $1-2$   & 8  & 0.027  & 1847 \\
    $2+$    & 34 & 0.13   & 4831 \\
    \hline
  \end{tabular}
  \label{tab:sample}
\end{table}

\begin{figure}
  \includegraphics[width=\columnwidth]{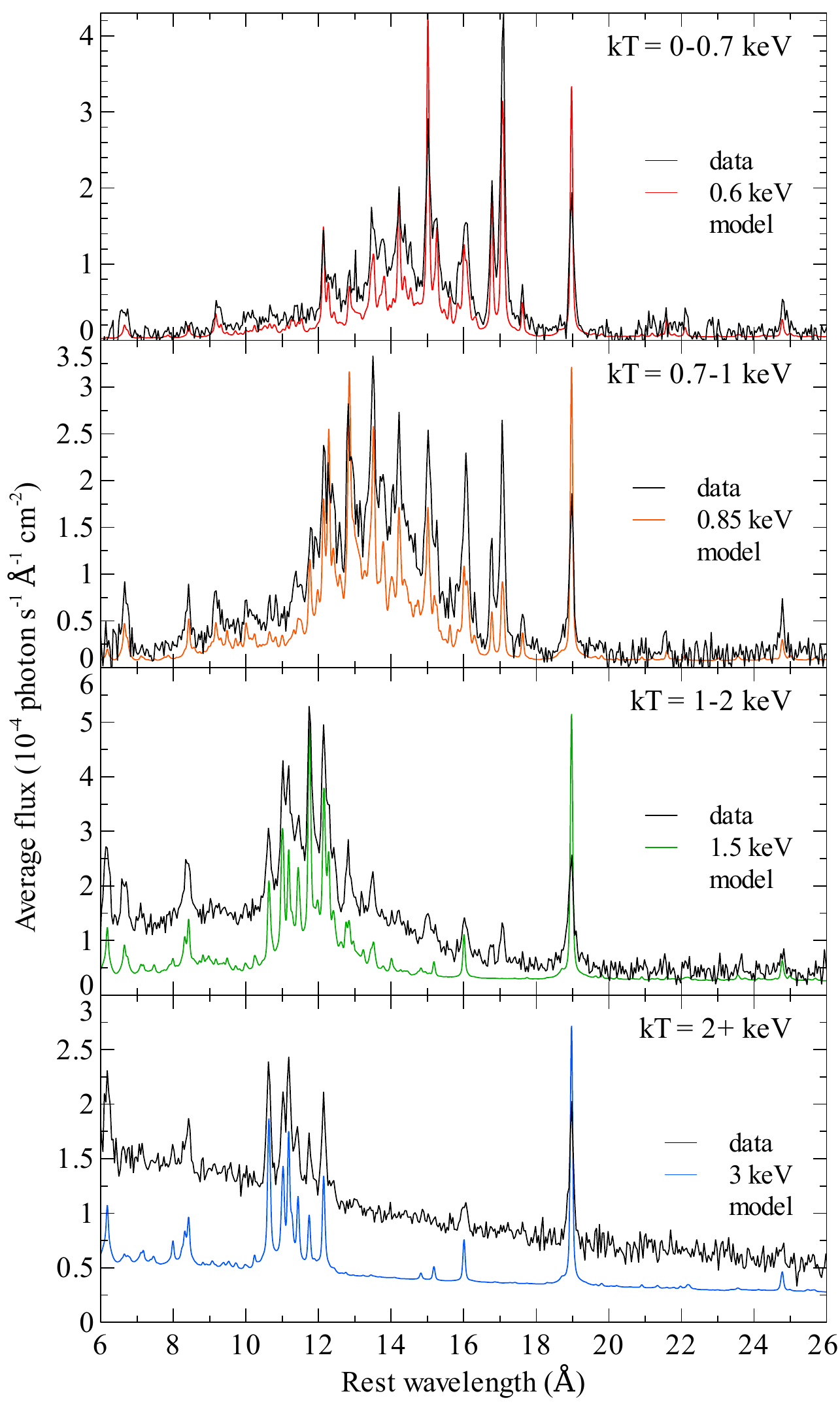}
  \caption{The mean spectra of objects divided into four temperature
    bins using the RGS-measured temperature. In each panel we plot
    model spectra with similar spectra resolution and temperature at
    Solar metallicity.}
  \label{fig:binT}
\end{figure}

We constructed stacked spectra for the clusters in four different
temperature bins using the average temperature measured from the RGS
spectra \citep[values are listed in][]{Sanders10_Broaden}. Table
\ref{tab:sample} shows details about the temperature bins. The average
spectra for the objects in the temperature bins are shown in
Fig~\ref{fig:binT} with Solar metallicity model spectra (computed
using \textsc{apec}; \citealt{SmithApec01}) at a representative
temperature and spectral resolution. The panels show that as the
temperature increases we see increasingly higher ionization states of
iron at shorter wavelengths. There are no obvious lines in the data
that are not accounted for in the model spectra. We see no evidence
for any lines that may come from, for example, dark matter decay
\citep[e.g.][]{Abazajian01}.

\begin{figure*}
  \includegraphics[width=\textwidth]{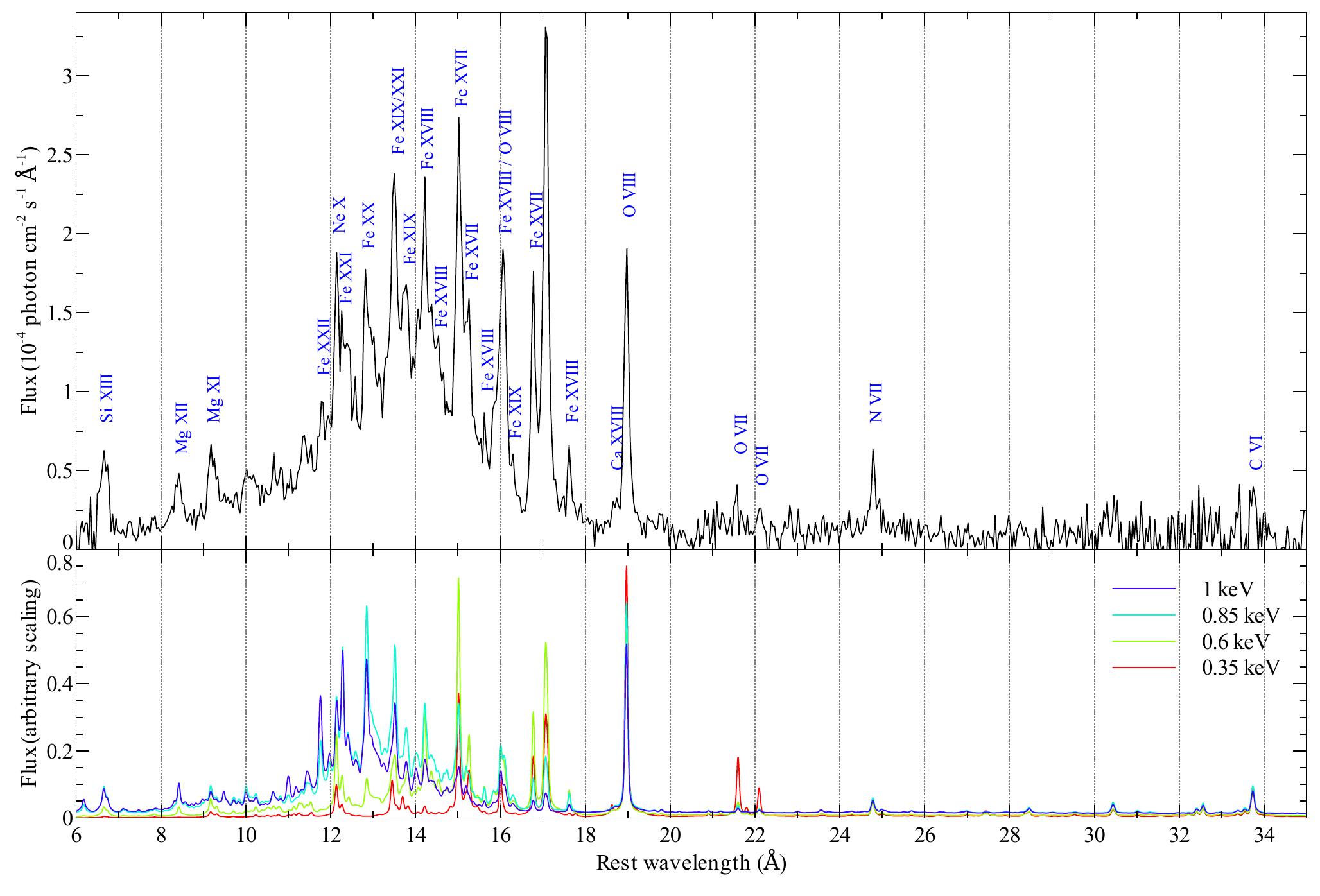}
  \caption{(Top panel) Mean of spectra of all objects below 1~keV
    temperature. This spectrum has a total \emph{XMM-Newton} exposure
    time of 1.2Ms. (Bottom panel) Model spectra with a range of
    temperature computed with \textsc{apec} at Solar metallicity.}
  \label{fig:coolobj}
\end{figure*}

The spectra for the cooler objects are line-rich. To examine the lines
in more detail we combine the two lower temperature bins. The mean
spectrum for the objects with temperatures less than 1~keV is shown in
Fig.~\ref{fig:coolobj}. On the plot we label the major lines and in
the lower panel show model spectra at a range of temperatures. In this
stacked spectrum we detect the recombination and forbidden lines of
the O\,\textsc{vii} triplet at 21.6 and 22.1{\AA},
respectively. O\,\textsc{vii} is a good indicator of material below
0.5~keV (Fig.~\ref{fig:linestrengths}). In addition we find the
C\,\textsc{vi} lines at 33.7{\AA}, although this part of the spectrum
is noisy and the continuum uncertain. C\,\textsc{vi} has been seen
previously in M87 \citep{Werner06M87}.

Due to the rebinning and fluxing process we cannot propagate the error
bars robustly from the original spectra. Instead, we used the standard
deviation of the data points in the spectral region around around the
O\,\textsc{vii} lines (20.8 to 24.5{\AA}, excluding the lines
themselves) to estimate the error bars on each spectral point. A
constant model fit over this region gives a $\chi^2$ of 108.4 with 92
degrees of freedom. If we add to the model two Gaussian components of
the same width at the wavelengths of the O\,\textsc{vii} lines, the
$\chi^2$ reduces to 87.4 with 89 degrees of freedom (a reduced
$\chi^2$ of 0.98). An F-test suggests that the probability of of this
improvement in fit happening by chance is $2.5\times 10^{-4}$. The
separate probability for the 21.6{\AA} line is $1.4\times 10^{-3}$ and
$0.06$ for the 22.1{\AA} line.  Similarly, the probability of false
detection of the C\,\textsc{vi} line is 0.02.

\section{Discussion}
The detection of O\,\textsc{vii} emission shows that there is material
below $\sim0.5$~keV in the X-ray waveband, at least in objects with a
mean central temperature below 1 keV. The signal does not seem to be
stronger when the sample is split into two bins (Fig.~\ref{fig:binT}),
which implies that it is unlikely that there is a single dominating
object or that only the cooler objects show O\,\textsc{vii} emission.

The flux of the O\,\textsc{vii} recombination and forbidden lines for
the sub-1~keV temperature sample, as obtained from the Gaussian model
fit, are shown in Table \ref{tab:lines}. The Fe\,\textsc{xvii} lines
at 16.78 and 17.08{\AA} were also fitted and the flux calculated (the
second line is actually two unresolved lines at 17.055 and
17.1{\AA}). For this fit, we fixed the continuum at the value obtained
from the O\,\textsc{vii} lines due to the proximity of neighbouring
lines. We did not measure the fluxes for the Fe\,\textsc{xvii} lines
around 15{\AA} because it is difficult to separate the continuum and
these lines.

\begin{table}
  \caption{Line fluxes observed in the mean sub-1~keV temperature spectrum compared
    to cooling flow models (units
    are $10^{-5}$~photon~$\cm^{-2} \s^{-1}$). Model fluxes are for a
    $1\Msunpyr$ cooling flow, cooling from 1.5~keV at Solar
    metallicity at a redshift of 0.01, absorbed with a Galactic column
    density of $10^{20}\psqcm$.}
  \begin{tabular}{llll}
    \hline
    Line                   & Observed        & Model (APEC) & Model (SPEX) \\
    \hline
    Fe \textsc{xvii} 16.78 & $2.52 \pm 0.07$ & 1.1          & 1.8 \\  
    Fe \textsc{xvii} 17.08 & $5.39 \pm 0.08$ & 2.5          & 4.1 \\
    O \textsc{vii} 21.6    & $0.31 \pm 0.09$ & 1.2          & 1.0 \\
    O \textsc{vii} 22.1    & $0.20 \pm 0.08$ & 0.63         & 0.67\\
    \hline
  \end{tabular}
  \label{tab:lines}
\end{table}

We calculated a model spectrum where gas is cooling isobarically at a
rate of $1\Msunpyr$ at a redshift of 0.01, from a temperature of
1.5~keV to 0.08~keV (assuming $H_0=70\kmpspMpc$). The spectrum is
absorbed photoelectrically at an equivalent Hydrogen column of
$10^{20}\pcmsq$, for both the \textsc{apec} and \textsc{spex} models.
The flux values for the same lines are shown in
Table~\ref{tab:lines}. For the \textsc{apec} model the Fe
\textsc{xvii} to O \textsc{vii} flux ratio is 2.0 (summing the pairs
of lines) and 3.5 for the \textsc{spex} model. In our average spectrum
the ratio is 16. Therefore if the Fe\,\textsc{xvii} and
O\,\textsc{vii} emission is caused by cooling, assuming Solar
metallicity ratios and no extra absorption on the coolest gas, there
is $1/4$ to $1/8$ of the emission below 0.5~keV temperature than
expected from the emission above those temperatures.

There may be more material cooling to a low temperatures if the cooler
gas is surrounded by absorbing material. The edge for oxygen
absorption occurs over a wavelength range 22.6 to 23.7{\AA} (seen in
the \textsc{tbnew} absorption model of \citealt{Wilms00}). This edge
is at longer wavelengths than the O\,\textsc{vii} triplet. If the
oxygen metallicity is low in the cooling regions, this may lead to
weak O\,\textsc{vii} emission. Oxygen lines are the main coolant in
the RGS waveband but iron emission is substantial at longer
wavelengths in the far UV. The abundances in the central
parts of these objects is likely to be unusual, particularly because
of depletion onto dust grains \citep[e.g.][]{Canning10}. Non-radiative
cooling of the ICM may also lead to there being less emission from
cooler temperatures than expected, for example, because of the mixing
of X-ray emitting material with colder gas \citep{FabianCFlow02}.

The average metallicities can also be estimated for the sample. Note
from Fig.~\ref{fig:linestrengths} that the ratios between the
hydrogenic O\,\textsc{viii}, N\,\textsc{vii} and C\,\textsc{vi}
emissivities are the same above a temperature of 0.25~keV. If there is
little material below this temperature (as is indicated by the lack of
O\,\textsc{vii} emission) then the ratios of these lines should be
good indicators of metallicity.  The Mg\,\textsc{xii} hydrogenic lines
around 8.4{\AA} could be used for abundance measurements above 1~keV
temperature but they are unfortunately blended with Fe\,\textsc{xxiv}.

\begin{table}
  \caption{Hydrogenic line fluxes observed in the mean $0-1$~keV
    spectrum compared to thermal models (units
    are $10^{-5}$~photon~$\cm^{-2} \s^{-1}$). Model fluxes are for a
    Solar metallicity plasma at 0.5 keV, absorbed by $10^{20}\psqcm$, with an
    Xspec normalization of $10^{-3}$ and at a redshift of 0.}
  \begin{tabular}{llll}
    \hline
    Line                    & Observed        & Model (APEC) & Model (SPEX) \\
    \hline
    O\,\textsc{viii} 19.0   & $2.62 \pm 0.08$ & 17           & 17  \\
    N\,\textsc{vii}  24.8   & $0.75 \pm 0.11$ & 1.2          & 1.4 \\
    C\,\textsc{vi}   33.7   & $0.40 \pm 0.14$ & 2.3          & 2.1 \\
    \hline
  \end{tabular}
  \label{tab:lines2}
\end{table}

Table \ref{tab:lines2} lists the line strengths for the hydrogenic
lines in the mean spectrum.  The error bars on the spectral data
points for the N\,\textsc{vii} and C\,\textsc{vi} lines were estimated
from the standard deviation of the data points around each line.  When
measuring the flux in the the O\,\textsc{viii} line we assumed the
same continuum level that we measured for the O\,\textsc{vii}
lines. We fitted separate continuum levels for N\,\textsc{vii} and
C\,\textsc{vi} (which are lower by 3 and 20 per cent from the
O\,\textsc{vii} continuum level, respectively).

We also tabulate in Table \ref{tab:lines2} the line strengths for a
simulated 0.5~keV spectrum at Solar metallicity, absorbed by a column
density of $10^{20}\psqcm$. The \textsc{apec} and \textsc{spex} models
show similar fluxes for the lines. At Solar metallicity there should
be a O/N flux ratio of $\sim 14$. We observe a ratio of 3.5, which
implies that the average nitrogen abundance relative to oxygen is $4.0
\pm 0.6\Zsun$. We note that the O\,\textsc{viii} emission line is
resonantly scattered which could lead to an underestimation of the O/N
ratio, although this effect is relatively small for an elliptical
galaxy like NGC~4636 \citep{Werner09}.

The N\,\textsc{vii} to O\,\textsc{viii} line ratio appears to be largest
in the 0.7-1 keV temperature range (Fig.~\ref{fig:binT}). The strength
of the N\,\textsc{vii} line relative to O\,\textsc{viii} decreases with
increasing temperature. This could either be due to real abundance
changes with temperature, or it may be due to the fact that we are
sampling larger regions for objects at higher redshifts.

We also applied spectral fitting to the combined spectrum for objects
with temperatures of less than 1~keV. Although there are difficulties
in interpreting the results from such a spectral fit, we confirmed
that nitrogen to oxygen ratio is greater than 3 times the solar
value. As the continuum at long wavelengths is uncertain, it is
unclear whether O/H metallicities are low or whether the N/H values
are high. The N/Fe ratio is around 1.7, although the Fe\textsc{xvii}
lines are poorly fit. The two temperature component model gives
temperatures of 0.34 and 0.71~keV, with the emission measure of the
lower temperature component 14 per cent of the emission measure of the
hotter component.

Nitrogen abundances appear to vary considerably in detailed
observations of individual objects, although the exact values are
dependent on modelling. \cite{Xu02} found that NGC~4636, a relatively
cool elliptical galaxy at $\sim 0.6$~keV, does not have strong
N\,\textsc{vii} emission, with N/Fe at approximately Solar
ratios. \cite{Werner09}, however, found $2.5\Zsun$ N/Fe in the core of
this object and high nitrogen abundances in four other elliptical
galaxies. In Centaurus the N/Fe ratio is around $3\Zsun$
\citep{SandersRGS08}. Centaurus has a mean RGS temperature of 1.6
keV. The N/Fe ratio in S\'ersic 159 is $0.1\Zsun$ \citep{dePlaa06},
with a mean 3 keV temperature. M87, which has a mean temperature of
1.8~keV shows a N/Fe ratio of $1.6\Zsun$ \citep{Werner06M87}. The N/Fe
ratios for Abell\,262, Abell\,3581 and HCG\,62 (relatively cool
objects with RGS temperatures less than 1.5 keV), are close to Solar
values \citep{SandersCool10}. It is unclear why there is such a wide
range of N/Fe values. A future study of a sample of bright objects
using the ratio of the N\,\textsc{vii} to O\,\textsc{viii} lines would
be less model dependent than using N/Fe.

Possible other observational evidence for high nitrogen abundances are
the high [N\,\textsc{ii}]$\lambda$6584/H$\alpha$ ratios in the nebulae
surrounding some central cluster galaxies, which can be as much as 3
\citep{Hatch07}. Ratios as high as this are difficult to achieve in
models without super-Solar nitrogen metallicities.

Examining the carbon metallicity, the C/O flux ratio implies an carbon
to oxygen abundance ratio of $0.85 \pm 0.32 \Zsun$.

Nitrogen and carbon metallicities are useful to measure because these
elements are not created by supernova explosions (unlike the other
elements detected in the X-ray waveband) and are typically created in
low to intermediate mass stars \citep[for a discussion
see][]{Molla06}.  Nitrogen can be made in a primary process in
intermediate mass stars, where oxygen and carbon generated before CNO
cycling is converted to nitrogen. This should give a nitrogen
enrichment which is independent of the initial metallicity of the
star. In addition, nitrogen can be created by a secondary process,
where carbon and oxygen present at the birth of the star is
transformed into nitrogen. This should give a nitrogen enrichment
proportional to the initial abundance. Oxygen enrichment is thought to
be mostly due to Type II supernovae, produced from massive stars.

The high average value of N/O observed in the sub-keV temperature
sample is at the extreme end of the distribution of H\,\textsc{ii}
regions and stars in late-type galaxies shown in
\cite{Henry00}. However, N/Fe has been found to be slightly supersolar
in a sample of early-type galaxies \citep{Toloba09}. The high N/O
values we observe may either be due to enhanced nitrogen abundances or
low oxygen abundances.

\section{Conclusions}
We examine the mean RGS spectrum of objects from the \emph{XMM-Newton}
archive in different temperature bins. For those objects with a
RGS-measured temperature of less than 1~keV we detect for the first
time O\,\textsc{vii} emission. This is indicative of material below
0.5~keV.

If the gas in the objects is cooling without any heating, there is
$1/4$ to $1/8$ of the expected O\,\textsc{vii} emission compared to the
16.8 and 17.1{\AA} Fe\,\textsc{xvii} flux.  This may imply that a small
fraction of the gas is cooling in these objects, the abundance ratios
of O and Fe are anomalous, or that there is absorbing material around
the coolest gas. An alternative explanation is non-radiative cooling
of the coolest gas.

The ratio of the hydrogenic lines in the spectrum are independent of
temperature, allowing estimates of the mean N/O and C/O ratios. In the
sample of objects below 1~keV the N/O ratio is $4.0 \pm 0.6
\Zsun$. The C/O ratio is $0.9 \pm 0.3 \Zsun$.

\section*{Acknowledgements}
ACF thanks the Royal Society for support.

\bibliographystyle{mnras}
\small
\bibliography{refs}

\end{document}